\documentstyle[multicol,aps,prl,floats]{revtex}

\begin{document}
\title{Passive advection in nonlinear medium\thanks{%
based on a talk delivered at ``International Conference on Turbulence:
Challenges for the 21st Century'' held at Los Alamos, May 18-21, 1998 and
dedicated to R.H. Kraichnan 70th birthday.}}
\author{Michael Chertkov}
\address{Physics Department, Princeton University, Princeton, NJ 08544}
\date{\today}
\maketitle

\begin{abstract}
Forced advection of passive tracer, $\theta $, in nonlinear relaxational
medium by large scale (Batchelor problem) incompressible velocity field at
scales less than the correlation length of the flow and larger than the
diffusion scale is considered. Effective theory explaining small scale
scalar fluctuations is proven to be linear, asymptotic free (downscales from
the scale of the pumping) and universal. Only three parameters are required
to decribe exhaustively the small scale statistics of scalar difference: two
velocity-dependent ones, average and dispersion ($\bar{\lambda}$ and $\Delta 
$ respectively) of the exponential stretching rate of a trial line element,
and $\alpha $, standing for average rate of linear damping of small scale
scalar fluctuations. $\alpha $ is an explicit functional of potential
chracterized medium nonlinearity and amplitude of $\theta ^{2}$ flux pumped
into the system. Structure functions show an extremely anomalous,
intermittent behavior: $\left\langle |\delta \theta _{r}|^{q}\right\rangle
\sim r^{\xi _{q}}$, $\xi _{q}=\min \left\{ q,\sqrt{\left[ \frac{\bar{\lambda}%
}{\Delta }\right] ^{2}+\frac{2\alpha q}{\Delta }}-\frac{\bar{\lambda}}{%
\Delta }\right\} $. No dissipative anomaly is found in the problem.
\end{abstract}

\draft

\section{Introduction}

Turbulence is very nonequilibrium state of nature, which becomes stationary
if energy is supplied permanently at large scales. To construct a theory of
turbulence means to describe temporal and spatial distributions of velocity
and of variety of different thermodynamic characteristics of the fluid, i.e.
density, if turbulence is compressible, temperature, if thermo-advection is
applied, relative concentration of components in case of multi-component
(color) flow, magnetic field distribution in a conducting fluid etc.
Dynamics of different fields describing a real turbulent flow is both
nonlocal and nonlinear. We call the general situation {\em active }to
emphasize the reciprocal character of interaction between velocity field and
thermodynamic characteristic(s). However, sometimes the effect of a
thermodynamic field on the velocity distribution is suppressed. It takes
place , for example, if scales are separated: a typical spatia-temporal
scale of velocity is much larger than one of a thermodynamic quantity. The
case, when it is theoretically justified to neglect the effect of back
reaction of thermodynamic field on velocity field in comparison with ones of
advection and nonlinearity is called {\em passive. }The passiveness does not
necessarily means linearity. Moreover, our objective is to study passive yet
nonlinear situation.

We consider dynamics of a thermodynamic quantity $\theta $, 
\begin{eqnarray}
\frac{d}{dt}\theta &=&-\frac{\delta H\{\theta \}}{\delta \theta }+\phi
\left( t;{\bf r}\right) ,  \label{Psieq} \\
H\{\theta \} &\equiv &\int d{\bf r}\left[ \frac{\kappa }{2}\left( \nabla
\theta \right) ^{2}+U(\theta )\right] ,  \label{H}
\end{eqnarray}
where $H\{\theta \}$ is a positive definite thermodynamic functional of the
system, $U(\theta )$ is confined ( $U\rightarrow +\infty $ at $\theta
\rightarrow \pm \infty $) potential, $\kappa $ is diffusion coefficient ; $%
\phi (t;{\bf r)}$ stands for statistically steady forcing providing a
constant supply of otherwise relaxational $\theta $-dynamics at large scales 
\footnote{%
We will discuss here the simplest case possible, when the thermodynamic
field is a scalar. Notice, however, that generalization of the discussed
theory for a vector or generally tensorial object is possible.}. $\theta $
is imbedded in a turbulent flow, i.e. the temporal derivative is extended by
sweeping term 
\begin{equation}
\frac{d}{dt}=\frac{\partial }{\partial t}+{\bf u}\nabla _{r},  \label{deriv}
\end{equation}
where incompressible velocity field ${\bf u(}t;r)$ is prescribed to be known
statistically.

We aim at finding the statistics of the passive scalar $\theta $ fixed by (%
\ref{Psieq}-\ref{deriv}) in the inertial interval of scales i.e. for scales
that are less than both the velocity correlation scale, $L_{u}$, and the
scale of the scalar supply, $L$, and larger than the diffusion scale.
Incompressible velocity field at those scales is modeled by the first term
of its local expansion in the radius vector connecting a reference point
with the current one, 
\begin{equation}
{\bf u}(t;{\bf r})=\hat{\sigma}(t){\bf r,}  \label{Batch}
\end{equation}
where $\hat{\sigma}(t)$ is $d\times d$ traceless random matrix of velocity's
derivatives.

The problem (\ref{Psieq}-\ref{Batch}) describes forced advection of a scalar
pollutant in the viscous-convective range absorbed or generated depending on
the sign of the nonlinear rate $\partial _{\theta }^{2}U(\theta )$, for
example via a chemical reaction with other species presented in abundance in
the flow. The problem is of fundamental importance for geophysical
atmospheric turbulence (see \cite{96Hay} for review). Other relevant
phenomenon is turbulent thermo-advection in a cell attached to thermal bath
(see \cite{91Mik} and reference therein). Then, $\partial _{\theta
}^{2}U(\theta )$ is nonlinear heat transfer coefficient and $\theta (t;{\bf r%
})$ measures local deviation from the bath temperature. Many regimes of
premixed turbulent combustion are also governed by (\ref{Psieq}-\ref{Batch}) 
\cite{88Wil},. The last (but not the least) situation to mention is a phase
ordering in a system described by scalar nonconserved order parameter (a
very well known object of the phase transition theory, see \cite
{Ma,77HH,82PP,93CH} for reviews) advected by large scale turbulent flow (see
also \cite{98Pat} on some discussion of reciprocal description of advection
and critical dynamics).

Our consideration will be based essentially on understanding, results and
general terminology emerged from studies of the pure problem of passive
scalar advection (no medium effect at all, $U=0$) having an almost five
decades of history (see Obukhov and Corrsin papers \cite{49Obu}, \cite{51Cor}
for the earliest contributions). Batchelor \cite{59Bat} has pioneered study
of the smooth velocity field limit (\ref{Batch}), which nowadays has grown
to be (through important contributions of many people \cite
{68Kra,69Coo,70Ors,74Kra,94SS,95CFKLa,94CGK,95BCKL,97BGK,98CFK,98KLS}) one
of the most advanced theory in the field . Temporal short-correlated but
spatially non-smooth model of velocity, one which gave more than two decades
later the first ever analytical evidence of intermittency in turbulence was
invented by Kraichnan. Structure functions of scalar difference in the
convective range, 
\begin{equation}
S_{q}(r)=\left\langle \left| \theta (t;r)-\theta (t;0)\right|
^{q}\right\rangle \sim r^{\xi _{q}},  \label{S2ndef}
\end{equation}
became the key object in the intermittency study. The anomalous scaling, $%
\Delta _{2n}\equiv n\xi _{2}-\xi _{2n}$, describing the law of the algebraic
growth with $L/r$ of the dimensionless ratio, $S_{2n}(r)/\left[
S_{2}(r)\right] ^{n}$, was shown to exist generically \cite
{95CFKLb,95GK,95SS}. The anomalous exponents were calculated perturbatively
in expansions about three non-anomalous ( $\Delta _{2n}=0$) limits, of large
space dimensionality $d$ \cite{95CFKLb,96CF}, of extremely non-smooth \cite
{95GK,96BGK} and almost smooth \cite{95SS} velocities respectively. A strong
anomalous scaling (saturation of $\xi _{2n}$ to a constant) was found for
the Kraichnan model at the largest $n$ by a steepest descent formalism \cite
{97Che,98BL}. Although the restricted asymptotic information about anomalous
exponents in the model is available a future possibility to establish
rigorously complete dependence of $\xi _{2n}$ on $n$, $d$ and degree of
velocity non-smoothness seems very unlikely (in a sense, recent Lagrangian
numerics \cite{98FMV} compensates the lack of rigorous information).

The problem (\ref{Psieq}-\ref{Batch}) discussed in the present paper is also
showing anomalous scaling, $\xi _{2n}<n\xi _{2}$, which is resolved
analytically for any kind of temporal correlations in $\hat{\sigma}$ and an
arbitrary (yet confined at $\theta \rightarrow \pm \infty $) potential $%
U(\theta )$. An important circumstance helping to find the general answer is
scale separation between scalar fluctuations at the integral scale, $L$, and
the current one, $r$, from convective range. For general $U(\theta )$, $%
\delta \theta _{r}$ obeys the same statistics as one would expect from an
auxiliary (linear!) problem with quadratic potential, $U^{*}(\theta )=\alpha
\theta ^{2}$, where $\alpha $ is given by average of $\partial _{\theta
}^{2}U\left( \theta \right) $ with respect to single point scalar
distribution, ${\cal P}_{1}\sim \exp \left[ -U(\theta )/\chi _{0}\right] $,
with $\chi _{0}\equiv \int_{0}^{\infty }dt\left\langle \phi (t;0)\phi
(0;0)\right\rangle $. $\alpha $ is always positive, i.e. at the smallest
scales the effect of nonlinearity, generally alternating between damping and
acceleration, is reduced to a pure linear damping. Linear problem, by
itself, happens to be solvable not only in the case of the short-correlated
velocity \cite{98Che} but for $\hat{\sigma}(t)$ statistics of a general
position (providing correlation functions of $\hat{\sigma}(t)$ taken at
different moments of time decays with the time shift faster than
algebraically). Finally, we have found anomalous exponents 
\begin{equation}
\xi _{q}=\min \left\{ q,\sqrt{\left[ \frac{\bar{\lambda}}{\Delta }\right]
^{2}+\frac{2\alpha q}{\Delta }}-\frac{\bar{\lambda}}{\Delta }\right\} ,
\label{zet2n}
\end{equation}
where $\bar{\lambda}$ and $\Delta $ are respectively average and dispersion
(with respect to $\hat{\sigma}$ averaging) of the rate of line stretching, $%
\lambda (t)=t^{-1}\ln [R(t)/R(0)]$ , with ${\bf R}(t)$, satisfied to $\dot{%
{\bf R}}(t)=\hat{\sigma}(t){\bf R}(t)$. To make the statement we used the
method of \cite{95CFKLa,94CGK}.

The anomalous behavior in the model under consideration differs from one
perceived in the Kraichnan model. First of all, $\xi _{2n}$ as a function of 
$n$ does not saturate to a constant at the largest $n$ but keeps growing
with $n$ as $\sqrt{n}$. Second ( and major) difference is associated with
the concept of dissipative anomaly. It is generally accepted to talk about
dissipative anomaly if some object calculated at zero dissipation ( $\kappa
=0$) does not coincide with its $\kappa \rightarrow 0$ counterpart. In the
Kraichnan model the anomalous scaling coexists with dissipative anomaly \cite
{94Kra}. However, the nonlinear problem (\ref{Psieq}-\ref{Batch}), as well
as its linear descendant, shows no dissipative anomaly while the anomalous
scaling is present. We base the important conclusion on the following {\em %
no anomaly criterium} (which, we believe, is general):{\em \ }if zero
dissipation analysis produces normalizable and everywhere positive solution
for the Probability Density Functional (PDF) of fluctuated field ( $\theta $
in our case) then the dissipative anomaly is absent.

The problem is formulated in Section II. To describe the scalar fluctuations
at a current scale from convective range we show how to integrate out the
large scale contribution in Section III. The scale separation results in
suppression of nonlinearity. The effective small scale theory appears to be
a linear one with uniform damping. All the final answers emerging from study
of the linear problem are presented in Section IV. The last Section V is
reserved for conclusions.

\section{Formulation of the problem}

(\ref{Psieq}-\ref{deriv}) describe advection of a passive scalar $\theta (t;%
{\bf r})$ by the smooth incompressible velocity field (\ref{Batch}). The
scalar is forced by random field $\phi (t;{\bf r})$, which for a sake of
simplicity is considered to be Gaussian thus fixed unambiguously by 
\begin{equation}
\left\langle \phi (t_{1};{\bf r}_{1})\phi \left( t_{2};{\bf r}_{2}\right)
\right\rangle =\chi (|{\bf r}_{1}-{\bf r}_{2}|)\delta \left(
t_{1}-t_{2}\right) ,  \label{phiphi}
\end{equation}
where the function $\chi (r)$ decays fast enough if $r$ exceeds the integral
scale $L\lesssim L_{u}$. $\chi _{0}=\chi (0)$ is the flux of $\theta ^{2}$
pumped into the system. $\hat{\sigma}$ is a random in time matrix process
described by its PDF, $\Phi \{\hat{\sigma}(t)\},$which is supposed to be
known. Diffusion is supposed to be small, such that the range of scales in
between $r_{d}=\sqrt{\kappa }/[S/\tau ]^{1/4}$ ( $S$ and $\tau $ are typical
values of the strain and velocity correlaion time respectively) and $L$ is
sufficiently large, $L/r_{d}\gg 1$.

We will be mainly aiming to find the two-point scalar PDF, 
\begin{equation}
{\cal P}_{2}\left( x_{+},x_{-}|r\right) \equiv \left\langle \delta \left(
x_{-}-\theta (t;0)+\theta (t;{\bf r})\right) \delta \left( x_{+}-\theta
(t;0)-\theta (t;{\bf r})\right) \right\rangle ,  \label{P2}
\end{equation}
and the scalar structure functions, 
\begin{equation}
{\cal S}_{2n}(r)\equiv \left\langle \left[ \theta (t;{\bf r})-\theta
(t;0)\right] ^{2n}\right\rangle ,  \label{S2n}
\end{equation}
where averaging, $\left\langle \cdots \right\rangle $, with respect to both $%
\hat{\sigma}(t)$ and $\phi (t;{\bf r})$, is assumed.

Other important objects used in the course of the forthcoming calculations
will be the two point scalar PDF, measured for particular $\hat{\sigma}(t)$
configuration , 
\begin{equation}
{\cal G}_{2}\left( x_{1},x_{2}|{\bf r}_{1,2};t;\{\hat{\sigma}(t^{\prime
});-\infty \leq t^{\prime }\leq t\}\right) \equiv \left\langle \delta \left(
x_{1}-\theta (t;{\bf r}_{1})\right) \delta \left( x_{2}-\theta (t;{\bf r}%
_{2})\right) \right\rangle _{\phi },  \label{P2sig}
\end{equation}
and the single point scalar PDF 
\begin{equation}
{\cal P}_{1}\left( x\right) \equiv \left\langle \delta \left( x_{-}-\theta
(t;{\bf r})\right) \right\rangle .  \label{P1}
\end{equation}
Notice, that the last object does not depend on the velocity field
statistics because of spatial homogeneity assumed.

Deep inside the convective range (at $L\gg r_{12}$), $|\theta _{1}-\theta
_{2}|\ll |\theta _{1}+\theta _{2}|$, and (\ref{P2sig}) can be decomposed
into the product 
\begin{equation}
{\cal G}_{2}\left( \theta _{1},\theta _{2}\left| {\bf r}_{1,2};t;\{\hat{%
\sigma}(t^{\prime });-\infty \leq t^{\prime }\leq t\}\right. \right) ={\cal P%
}_{1}(\theta _{1})*{\cal G}_{-}\left( \theta _{1}-\theta _{2}\left| {\bf r}%
_{1}-{\bf r}_{2};t;\{\hat{\sigma}(t^{\prime });-\infty \leq t^{\prime }\leq
t\}\right. \right) .  \label{decomp}
\end{equation}
The average of (\ref{decomp}) over $\hat{\sigma}$ reads as 
\begin{equation}
{\cal P}_{2}\left( \theta _{1},\theta _{2}\left| r_{12}\right. \right) =%
{\cal P}_{1}(\theta _{1})*{\cal P}_{-}\left( \theta _{1}-\theta _{2}\left|
r_{12}\right. \right) .  \label{decompav}
\end{equation}

The assumption on {\em the absence of the dissipative anomaly }in the case
of a very small diffusion lies in the core of our consideration. The formal
consequence of the statement is the possibility to drop off the dissipative $%
\kappa $-dependent term from (\ref{H}) already on the dynamical (yet
unaveraged) level. The no-anomaly assumption will be justified by the
positivity and normalizability of the derived answers for PDFs.

\section{Reduction of the nonlinear problem to a linear one}

In the absence of diffusion (\ref{Psieq}-\ref{Batch}) can be integrated
along the Lagrangian trajectories (characteristics) 
\begin{eqnarray}
\frac{d}{dt^{\prime }}\theta (t^{\prime };{\bf \rho (}t^{\prime })) &=&-%
\frac{dU}{d\theta }\left| _{\theta (t^{\prime };{\bf \rho (}t^{\prime
}))}\right. +\phi \left( t^{\prime };{\bf \rho }(t^{\prime })\right) ,
\label{Lagr} \\
\frac{d}{dt^{\prime }}{\bf \rho }(t^{\prime }) &=&\hat{\sigma}(t^{\prime })%
{\bf \rho }(t^{\prime }),\hspace{0.2in}{\bf \rho }(t)={\bf r,\hspace{0.2in}}%
-\infty <t^{\prime }<t.  \label{Lagrrho}
\end{eqnarray}
Notice, that the nonlinearity leads to scalar generation in the region of
convex ($\partial _{\theta }U>0$) potential while it dumps scalar
fluctuations if $\partial _{\theta }U<0$. Fokker-Planck equations (see \cite
{89CCK} for similar calculations) derived out of (\ref{Lagr},\ref{Lagrrho})
by means of direct averaging over the Gaussian noise, $\phi $, are 
\begin{eqnarray}
\left[ \partial _{\theta }\frac{dU\left( \theta \right) }{d\theta }-\chi
\left( 0\right) \partial _{\theta }^{2}\right] {\cal P}_{1} &=&0,
\label{eqP1} \\
\left[ \partial _{t}+\sum\limits_{i=1,2}\left( \sigma ^{\mu \nu
}(t)r_{i}^{\mu }\partial _{r_{i}}^{\nu }-\partial _{\theta _{i}}\frac{%
dU\left( \theta _{i}\right) }{d\theta _{i}}\right)
-\sum\limits_{i,j=1,2}\chi \left( r_{i}-r_{j}\right) \partial _{\theta
_{i}}\partial _{\theta j}\right] {\cal G}_{2} &=&0,  \label{eqG2}
\end{eqnarray}
where ${\cal G}_{2}$ is not stationary, since it does dependent on time
explicitly through $\hat{\sigma}(t)$. Integrating (\ref{eqG2}) with respect
to $\theta _{+}=\theta _{1}+\theta _{2}$ and assuming that the integral is
formed at $|\theta _{1}-\theta _{2}|\ll |\theta _{1}+\theta _{2}|$, where (%
\ref{decomp}) is valid, we arrive at the close equation for the scalar
difference PDF, 
\begin{equation}
\left\{ \partial _{t}+\left( \sigma ^{\mu \nu }(t)r^{\mu }\partial _{r}^{\nu
}-\alpha \partial _{x}x\right) -2\left[ \chi (0)-\chi \left( r\right)
\right] \partial _{x}^{2}\right\} {\cal G}_{-}\left( x\left| {\bf r};t;\{%
\hat{\sigma}(t^{\prime });-\infty \leq t^{\prime }\leq t\}\right. \right) =0,
\label{EqGmin}
\end{equation}
where $\alpha $ is defined as the following average over the large scale $%
\theta $ statistics 
\begin{equation}
\alpha \equiv \left\langle \frac{d^{2}U\left( \theta \right) }{d\theta ^{2}}%
\right\rangle _{LS}\equiv \int\limits_{-\infty }^{\infty }d\theta \frac{%
d^{2}U\left( \theta \right) }{d\theta ^{2}}{\cal P}_{1}\left( \theta \right)
.  \label{alpha}
\end{equation}
The normalized and everywhere positive solution of (\ref{eqP1}) is 
\begin{equation}
{\cal P}_{1}\left( \theta \right) =\frac{\exp \left[ -U(\theta )/\chi
_{0}\right] }{\int\limits_{-\infty }^{\infty }d\theta \exp \left[ -U(\theta
)/\chi _{0}\right] }.  \label{P1sol}
\end{equation}
Substitution of (\ref{P1sol}) into (\ref{alpha}) gives 
\begin{equation}
\alpha =\frac{\left\langle \left[ \frac{dU\left( \theta \right) }{d\theta }%
\right] ^{2}\right\rangle _{LS}}{\chi _{0}}=\frac{\int\limits_{-\infty
}^{\infty }d\theta \left[ \frac{dU\left( \theta \right) }{d\theta }\right]
^{2}\exp \left[ -U(\theta )/\chi _{0}\right] }{\chi _{0}\int\limits_{-\infty
}^{\infty }d\theta \exp \left[ -U(\theta )/\chi _{0}\right] },  \label{alpos}
\end{equation}
i.e. $\alpha $ is principally positive constant, does not matter what is the
form of the potential $U(\theta )$ be (provided it grows with $|\theta |$ at 
$\theta \rightarrow \pm \infty $). Therefore, we have found that at the
smallest scales regions of scalar generation are suppressed statistically.

On the basis of (\ref{eqP1}) and (\ref{EqGmin}) we conclude that, from the
point of view of the small scale statistics of scalar difference our problem
is equivalent to the linear one, with $dU(\theta )/d\theta $ being replaced
just by $\alpha \theta $. In other terms, we may proceed averaging the
linear dynamical equation 
\begin{equation}
\partial _{t}\theta +\sigma ^{\mu \nu }(t)r^{\mu }\nabla _{r}^{\nu }\theta
=-\alpha \theta +\phi (t;{\bf r}),  \label{thetalin}
\end{equation}
instead of the original nonlinear one. The steady distribution of the scalar
difference underlined (\ref{thetalin}) was the subject of the recent paper 
\cite{98Che}, the method and results of which will be briefed and
generalized (for the case of finite correlated velocity) in the next Section.

\section{Velocity averaging. Anomalous scaling.}

The linear analog of (\ref{Lagr}) is 
\begin{equation}
\theta (t;{\bf r})=\int\limits_{0}^{\infty }dt^{\prime }\exp \left[ -\alpha
t^{\prime }\right] \phi \left( t^{\prime };{\bf \rho }(t-t^{\prime })\right)
.  \label{Lagrlin}
\end{equation}

For the purpose of the $2n$-th structure function calculation it is utmost
enough to consider the simultaneous product, $F_{1\cdots 2n}\equiv
\left\langle \theta _{1}\cdots \theta _{2n}\right\rangle $, which is
according to (\ref{Lagrlin},\ref{Lagrrho}) is 
\begin{eqnarray}
F_{1\cdots 2n} &=&\sum\limits_{\{i_{1,\cdots },i_{2n}\}}^{\{1,\cdots
,2n\}}\left\langle \prod\limits_{k=1}^{n}\int\limits_{0}^{\infty
}dt_{k}e^{-\alpha t_{k}}\chi \left[ \hat{W}(t_{k})r_{i_{k};i_{k+1}}\right]
\right\rangle _{\hat{\sigma}},  \label{F12n} \\
\hat{W}(t) &\equiv &T\exp \left[ \int\limits_{0}^{t}dt^{\prime }\hat{\sigma}%
(t^{\prime })\right] ,\hspace{0.2in}\frac{d\hat{W}(t)}{dt}=\hat{\sigma}(t)%
\hat{W}(t).  \label{W}
\end{eqnarray}
Calculation of $F_{1\cdots 2n}$ is essentially simplified for the collinear
configuration, ${\bf r}_{i}={\bf n}$ $r_{i}$, when the $2n\times (d-1)$
parametric average (\ref{F12n}) is reduced to the following
single-parametric one 
\begin{equation}
F_{1\cdots 2n}=\sum\limits_{\{i_{1,\cdots },i_{2n}\}}^{\{1,\cdots
,2n\}}\left\langle \prod\limits_{k=1}^{n}\int\limits_{0}^{\infty
}dt_{k}e^{-\alpha t_{k}}\chi \left[ e^{\eta (t_{k})}r_{i_{k};i_{k+1}}\right]
\right\rangle ,  \label{Fq2nsc}
\end{equation}
with $r_{ij}\equiv |{\bf r}_{i}-{\bf r}_{j}|$. Here, the longitudinal
stretching rate, $\eta (t)\equiv \ln |\hat{W}(t){\bf n}|,$ is the only
fluctuating quantity left. The $\alpha =0$ version of (\ref{Fq2nsc}) was
calculated in \cite{95CFKLa} for the $d=2$ case and generalized for any $%
d\geq 2$ in \cite{94CGK} via a change of variables and further
straightforward transformation of the path integral standing for the average
over $\hat{\sigma}(t)$. It is shown \cite{95CFKLa,94CGK} that at the largest
times $\eta $-measure is a shifted Gaussian one, 
\begin{equation}
{\cal D}\eta (t)\exp \left[ -\int\limits_{0}^{\infty }dt\frac{\left( \dot{%
\eta}-\bar{\lambda}\right) ^{2}}{2\Delta }\right] ,  \label{Meseta}
\end{equation}
characterized by two parameters only: average, $\bar{\lambda}$, and
dispersion, $\Delta $, of the Lyapunov exponent, $\int\limits_{0}^{t}dt^{%
\prime }\eta (t^{\prime })/t$. (\ref{Meseta}) applied to (\ref{Fq2nsc})
produces 
\begin{equation}
\frac{F_{1\cdots 2n}}{n!}=\int \left[ \prod\limits_{i=1}^{n}dt_{i}d\eta
_{i}\right] \exp \left[ \frac{\bar{\lambda}}{\Delta }\eta _{1}\!-\!\frac{%
\bar{\lambda}^{2}}{2\Delta }t_{1}\right] \sum\limits_{\{k_{i},\cdots
,k_{2n}\}}^{\{1,\cdots ,2n\}}\prod_{i=1}^{n}\left[ e^{2\alpha t_{i}}\chi
\left( e^{\eta _{i}}r_{k_{2i},k_{2i+1}}\right) G\left( t_{i-1,i};\eta
_{i-1,i}\right) \right] ,  \label{F12n2}
\end{equation}
where $\eta _{i}$ ( $i\leq n$) integrations are not restricted, $0\leq
t_{n}\leq \cdots \leq t_{1}\leq \infty $, $t_{n+1}=\eta _{n+1}=0$, $%
t_{i,k}\equiv t_{i}-t_{k}$ (with equivalent notations for $\eta $) and 
\begin{equation}
G\left( t;\eta \right) \equiv \frac{\exp \left[ -\frac{\eta ^{2}}{2\Delta t}%
\right] }{\sqrt{2\pi \Delta t}}.  \label{Green}
\end{equation}
Both $\bar{\lambda}$ and $\Delta $ are unambiguously fixed by $\Phi \{\hat{%
\sigma}(t)\}$. The integrand of (\ref{F12n2}) decays exponentially in time
with the major contribution into the integral formed at $t_{i}\sim 1/\alpha $%
. The leading term does not depend on any $r_{ij}$ and gives no contribution
into $2n$-th order structure function. The first actual $r$-dependent
contribution stems from $n-1$ temporal integrals formed at $\tau \sim
1/\alpha $, and one at $t_{i}\sim \tau _{r}\sim \ln \left[ L/r\right] /\max
\{\alpha ,D\}$. This special integration brings a spatial dependence into
the object, therefore on a single distance. Generally, there exists a
variety of terms with all the possible combinations, like term with $k$
integration formed at $\tau $, while $n-k$ ones at $\tau _{r}$, and
therefore dependent explicitly on $2(n-k)$ spatial points. However, we are
looking exclusively for a term dependent on all the $2n$ points since only
such a term contributes $S_{2n}(r)$. It is really simple to calculate the
scaling of this term making use of the temporal separation, $\tau _{r}\gg
\tau $. Indeed, the large time contribution may be extracted out of (\ref
{F12n2}) in a saddle-point calculation. Variation of all the exponential
terms in (\ref{F12n2}) with respect to $t_{i}$ gives a chain of saddle
equations. The $\chi $ functions in the integrand of (\ref{F12n2}) limits
the $\eta $ integrations from above by $\ln \left[ L/r\right] .$ Therefore,
the desirable $2n$-points contribution forms at $t_{i}=\sqrt{\bar{\lambda}%
/\left[ \Delta \left( 2\alpha n\Delta +\bar{\lambda}^{2}/2\right) \right] }%
\ln \left[ L/r\right] $, and $\eta _{i}=\ln \left[ L/r\right] $, where it is
assumed that in the leading logarithmic order there is no need to
distinguish between contributions of different separations $r_{ij}$.
Substituting the saddle-point values of $t_{i}$ and $\eta _{i}$ into (\ref
{F12n2}) one arrives at the anomalous part of (\ref{zet2n}), with $q=2n$.
The normal-scaling counterpart of (\ref{zet2n}) originates from expansion of 
$\chi (r)$ from (\ref{F12n2}) in a regular series in $r^{2}$.

The basic physics of nonzero $\xi _{2n}$ (means deviating from the naive
balance of pumping and advection) and generally anomalous ($\xi _{2n}<n\xi
_{2}$) scaling at $\alpha >0$ can be stated quite clearly. According to (\ref
{Lagrlin}) the advection changes scales but not amplitude, while the
amplitude of injected scalar field decays exponentially from the time of
injection at the constant rate $\alpha $. The temporal integrals in (\ref
{F12n2}) forms at the mean time to reach a scale which is proportional to
the negative log of the scale. However, the effective spread in the factor
by which amplitude has decayed, upon reaching a given scale, increases as
scale decreases. It is why $\xi _{2n}>0$. Also there is more room for
fluctuations about the mean time due to the interference between the
exponential decay of the scalar amplitude and fluctuations of the stretching
rate $\eta $. Thus intermittency increases with a scale size decrease.

Another way to derive ( \ref{zet2n}) out of (\ref{Lagrlin}) is to construct $%
S_{2n}(r)$ directly. It is easy to check that the structure functions of
different orders are produced by the PDF satisfied to 
\begin{equation}
\bar{\lambda}r^{1-2\bar{\lambda}/\Delta }\partial _{r}r^{1+2\bar{\lambda}%
/\Delta }\partial _{r}{\cal P}_{-}+\alpha \partial _{x}\left( x{\cal P}%
_{-}\right) +\left[ \chi _{0}-\chi \left( r\right) \right] r^{2}\partial
_{x}^{2}{\cal P}_{-}=0,  \label{Pmin}
\end{equation}
The solution of (\ref{Pmin}), in the regime where you can neglect the $\chi $%
-dependent term is 
\begin{equation}
{\cal P}_{-}\left( x\left| r\right. \right) =\frac{1}{2\pi i}\frac{1}{\theta
_{L}}\int\limits_{0^{+}-i\infty }^{0^{+}+i\infty }ds\left[ \frac{\theta _{L}%
}{x}\right] ^{s+1}\left[ \frac{r}{L}\right] ^{\sqrt{d^{2}/4+\alpha s/[D(d-1)]%
}-d/2}a_{s}.  \label{Pminans}
\end{equation}
Here, $a_{s}$ is a function fixed by matching at the integral scale,
roughly, ${\cal P}_{-}\left( x\left| L\right. \right) \sim {\cal P}%
_{1}\left( x\right) $, where ${\cal P}_{1}\left( x\right) $ is given by (\ref
{P1sol}). The PDF (\ref{Pminans}) is positive and normalizable, that,
therefore, confirms the initial hypothesis on the absence of dissipative
anomaly. Also, (\ref{Pminans}) shows that (\ref{zet2n}) holds for general
(not only even integer) positive $q$.

\section{Conclusion}

We have shown that the nonlinear problem (\ref{Psieq}-\ref{Batch}) is
reduced to a linear one at the smallest (still from convective, not
dissipative, range) scales. The asymptotic theory remembers about initial
nonlinearity through the effective damping coefficient (\ref{alpha}). The
linear problem was solved for the general case of arbitrary correlated in
time large scale velocity field.

The most important feature of the problem appears to be the absence of
dissipative anomaly, the point which was guessed initially. Selfconsistency
of the hypothesis was confirmed afterwards by checking positivity and
normalizability of the final expression (\ref{Pminans}) for PDF. Of course,
the absence of dissipative anomaly is, by no means, a common situation in
turbulence. It is however suggestive to start analyzing any new turbulent
problem from the simple ''no anomaly'' test.

We discussed only the Batchelor case of large scale velocity. The
simplification appears to be a very important both for the fact of absence
of dissipative anomaly and solvability of the problem. The Batchelor case is
very special, since the Lagrangian dynamics of $n$ particles, generally
described by $n(d-1)$ degrees of freedom, is reduced to dynamics of $d-1$
eigenvalues of stretching matrix. This lies in the core of the Batchelor
problem's solvability. Also, in the Batchelor case scaling dimension of eddy
diffusivity operator coincides with one of the $\alpha $ (damping) dependent
term. The coincidence of exponents explains the anomalous scaling,
particularly the continuous dependance of the exponents on $\alpha $. Any
multiscale velocity field (say taken from the Kraichnan model) leads, first,
to appearance of the dissipative anomaly already on the medium free ( $U=0$)
level, and second to dis-balance of the scaling dimensions of advective and
medium-originated contribution into the eddy-diffusivity operator, resulting
in complete screening of any medium effect in the convective range. We
conclude by this guess, which is rather brave (and , of course, is not
rigorous at all). More studies, first of all on the nature of dissipative
anomaly, are required in this direction.

\smallskip

I thank L. Kadanoff, I. Kolokolov, B. Meerson, A. Patashinski, R.
Pierrehumbert, A. Polyakov, B. Shraiman and V. Yakhot for inspiring
discussions. Valuable comments of P. Constantin, G. Falkovich, R. Kraichnan,
V. Lebedev, and M. Vergassola are greatly appreciated. The work was
supported by a R.H. Dicke fellowship.


\end{document}